# The study of propagation characteristics of millimeter-wave vortex in magnetized plasma by using FDTD Method

Chenxu Wang[1]*, Hideki Kawaguchi[2]*, Hiroaki Nakamura[1], Shin Kubo[3]

[1]*National Institute for Fusion Science, Toki, Gifu 509-5292, Japan.*
[2]*Muroran Institute of Technology, Muroran, Hokkaido 050-0071, Japan*
[3]*Chubu University, Kasugai, Aichi 487-8501, Japan*

*E-mail: naka-lab@nifs.ac.jp; kawa@muroran-it.ac.jp;

It is pointed out that millimeter-wave vortex may contribute an efficient plasma heating since it was found that the millimeter-wave vortex can propagate in magnetized plasma even in which the normal plane wave is in cut-off condition. Then, it was assumed that the vortex field was the Laguerre-Gaussian (L-G) mode which is free-space solution, but the generation and stable propagation of the L-G mode vortex are not easy in the millimeter frequency range. On the other hand, it is known that millimeter-wave hybrid mode of cylindrical corrugated waveguide has also vortex property. In this paper, we investigate propagation characteristics of millimeter-wave vortex of a hybrid mode of cylindrical corrugated waveguide in the magnetized plasma by using three dimensional numerical simulations with finite-difference time-domain (FDTD) method.

## 1. Introduction

Since it was pointed out by L.Allen in 1991 that[1] the optical vortex, which has a helical wave-front, carries the orbital angular momentum (OAM) for the Laguerre-Gaussian beam, various investigations on applications of the optical vortex have been carried out, for example, large capacity fiber communication[2], optical tweezer[3] and so on. Although most of studies on such the vortex field are for optical frequency, it is known that such the phenomena of the vortex fields exist in other frequency ranges of X-rays, ultraviolet, microwave, and millimeter wave[4-10] as well. Then, from a view point of plasma science, it is one of the most remarkable contributions related to the vortex phenomena that possibilities of stable propagation of millimeter-wave vortex were shown in the magnetized plasma region even in which the standard plane wave is in cut-off condition[11]. This means that a more efficient method for plasma heating can be provided by using the millimeter-wave vortex than that of conventional plane wave. However, in the discussion of the stable propagation of the vortex fields in the magnetized plasma[11], the millimeter-wave vortex was assumed to be Laguerre-Gaussian (L-G) mode vortex, which is difficult to be generated and propagated stably in millimeter-wave frequency. On the other hand, it was pointed out that there exist stable hybrid modes vortex in the cylindrical corrugated waveguide, which carry well-defined orbital angular momentum[12]. In this paper, we discuss the propagation characteristics of millimeter-wave vortex fields of the hybrid modes in the magnetized plasma by using full three-dimensional simulation of the finite-difference time-domain (FDTD) method.

## 2. Millimeter-wave vortex of corrugated waveguide

2.1 Millimeter-wave vortex of hybrid modes in corrugated waveguide

Both of L-G mode vortex and hybrid mode vortex carry well-defined orbital angular momentum. But the L-G mode vortex is a free-space solution, on the other hand, hybrid mode vortex is the solution in cylindrical corrugated waveguide (Fig.1). In particularly, the higher modes of the L-G mode and hybrid mode have different field distribution each other. And then, it is not easy to generate the L-G mode vortex in millimeter-wave frequency. Accordingly, we consider to use the hybrid mode generated in the cylindrical corrugated waveguide for millimeter-wave plasma heating.

Under the following asymptotic approximation for wave number $k = \omega/c$ and phase constant for $z$-direction $\beta$,

$$k^2 \cong \beta^2 \gg k^2 - \beta^2 = k_c^2, \tag{1}$$

there exist stable propagation ($z$-direction) solutions to the Helmholtz's equation for the cylindrical corrugated waveguide,[12]

$$E_x = -iE_{mn}\frac{\beta}{k_c}J_{m-1}(k_c\rho) \times e^{i(\omega t - \beta z \pm (m-1)\emptyset)}, \tag{2-1}$$

$$E_y = -iE_{mn}\frac{\beta}{k_c}J_{m-1}(k_c\rho) \times e^{i\left(\omega t - \beta z \pm (m-1)\emptyset \pm \frac{\pi}{2}\right)}, \tag{2-2}$$

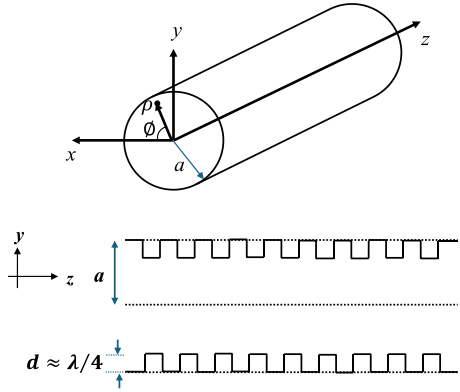

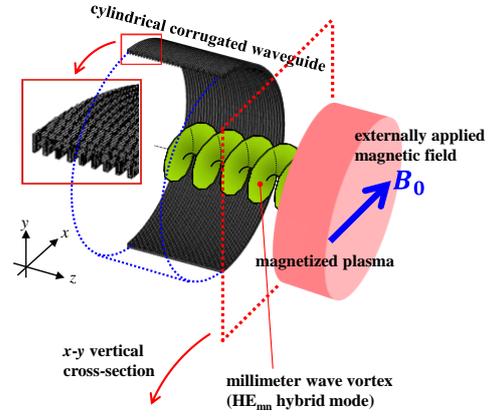

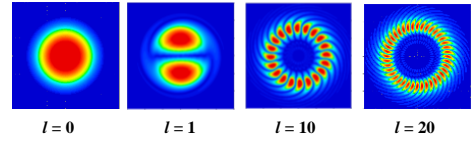

Fig.1 Cylindrical corrugated waveguide

Fig.2 Configuration of plasma heating by hybrid mode vortex

$$E_z = E_{mn}J_m(k_c\rho) \times e^{i(\omega t - \beta z \pm m\phi)}, \tag{2-3}$$

$$H_x = -iE_{mn}\frac{\beta}{Z_0 k_c}J_{m-1}(k_c\rho) \times e^{i\left(\omega t - \beta z \pm (m-1)\phi \pm \frac{\pi}{2}\right)}, \tag{2-4}$$

$$H_y = -iE_{mn}\frac{\beta}{Z_0 k_c}J_{m-1}(k_c\rho) \times e^{i(\omega t - \beta z \pm (m-1)\phi)}, \tag{2-5}$$

$$H_z = -iE_{mn}\frac{1}{Z_0}J_m(k_c\rho) \times e^{i(\omega t - \beta z \pm m\phi)}, \tag{2-6}$$

in the Cartesian coordinate $(x, y, z)$ shown in Fig.1, where $J_m(x)$ is the first kind of Bessel's function of order of $m$, and $n$ specifies zero point of the Bessel's function, $z_0 = \sqrt{(\mu_0/\varepsilon_0)}$ is wave impedance, $E_{mn}$ is amplitude of $E_z$. The cut-off wave number $k_c$ and inner radius of corrugated waveguide $a$ satisfies the following condition,

$$J_{m-1}(k_c a) = 0. \tag{3}$$

If we define power $P$ and angular momentum $\boldsymbol{M}$ of the hybrid mode as follows,

$$P = \frac{1}{2T}\int_0^T dt \int_S (\boldsymbol{E} \times \boldsymbol{H}^*) \cdot d\boldsymbol{S}, \tag{4}$$

$$\boldsymbol{M} = \frac{1}{2T}\int_0^T c\,dt \int_S \left(\boldsymbol{\rho} \times \frac{1}{c^2}(\boldsymbol{E} \times \boldsymbol{H}^*)\right) \cdot d\boldsymbol{S}, \tag{5}$$

we can find the following relation for $P$ and $M_z$,

$$\frac{M_z}{P} = \frac{\pm m}{\omega} = \frac{\pm l + \sigma_z}{\omega}, \text{ where } l=m-1,\ \sigma_z = +1,\ \text{or } -1. \tag{6}$$

where $\boldsymbol{\rho}$ is a relative position vector from $z$-axis in $x$-$y$ cross-section, $T$ is the time period.

This relation can be interpreted that $l$ and $\sigma_z$ correspond to orbital and spin angular momentum, respectively, according to the discussion by Allen[1),12)]. It is notified that the value of the topological charge $l$ of the vortex fields of the hybrid mode is ($m$-1). We here consider that the hybrid mode millimeter-wave vortex, which is obtained from the cylindrical corrugated waveguide, is illuminated to the magnetized plasma (see Fig.2). Examples of distributions of electric field intensity of hybrid mode vortex fields in *x-y* vertical cross-section are depicted in Fig.2.

2.2 Macro-model of magnetized plasma for propagation of millimeter-wave vortex

Since coupling analysis of electromagnetic wave and full simulation of plasma particles need extremely large computer costs, we here employ the following Drude-Lorentz macro-model for behavior of the magnetized plasma by using electron displacement density vector $\boldsymbol{P}$ and current density vector $\boldsymbol{J} = d\boldsymbol{P}/dt$,

$$\frac{d\boldsymbol{J}}{dt} + \gamma \boldsymbol{J} + \omega_0^2 \boldsymbol{P} = \varepsilon_0 \omega_p^2 \left(\boldsymbol{E} + \frac{1}{n_e q_e}\boldsymbol{J} \times \boldsymbol{B}_0\right), \tag{7}$$

where $\gamma$ and $n_e$ is dumping coefficient, electron density, respectively, $\omega_p = \sqrt{q_e^2 n_e / \varepsilon_0 m_e}$ is a plasma angular frequency, $\boldsymbol{E}$ is electric field vector and $\boldsymbol{B}_0$ is externally applied magnetic field.

# 3. FDTD formulation for millimeter-wave vortex propagation in magnetized plasma

FDTD analysis of the propagation of the millimeter-wave vortex in the magnetized plasma are carried out by coupling of discretized Maxwell's equations in 3D grid space for electric and magnetic fields $\boldsymbol{E}$, $\boldsymbol{H}$,

$$\boldsymbol{E}^{n+1} = \frac{\frac{\varepsilon_0}{\Delta t} - \frac{\sigma}{2}}{\frac{\varepsilon_0}{\Delta t} + \frac{\sigma}{2}} \boldsymbol{E}^n + \frac{1}{\frac{\varepsilon_0}{\Delta t} + \frac{\sigma}{2}} \nabla \times \boldsymbol{H}^{n+\frac{1}{2}} - \frac{1}{\frac{\varepsilon_0}{\Delta t} + \frac{\sigma}{2}} \boldsymbol{J}^{n+\frac{1}{2}}, \tag{8}$$

$$\boldsymbol{H}^{n+\frac{1}{2}} = \boldsymbol{H}^{n-\frac{1}{2}} - \frac{\Delta t}{\mu_0} \nabla \times \boldsymbol{E}^n, \tag{9}$$

and the magnetized plasma macro-model (7) for $\boldsymbol{P}$ and $\boldsymbol{J}$,

$$\boldsymbol{P}^{n+1} = \Delta t \boldsymbol{J}^{n+\frac{1}{2}} + \boldsymbol{P}^n, \tag{10}$$

$$\frac{\boldsymbol{J}^{n+\frac{1}{2}} - \boldsymbol{J}^{n-\frac{1}{2}}}{\Delta t} + \gamma \frac{\boldsymbol{J}^{n+\frac{1}{2}} + \boldsymbol{J}^{n-\frac{1}{2}}}{2} + \omega_0^2 \boldsymbol{P}^n = \varepsilon_0 \omega_p^2 \boldsymbol{E}^n + \frac{\varepsilon_0 \omega_p^2}{q_e n_e} \frac{\boldsymbol{J}^{n+\frac{1}{2}} + \boldsymbol{J}^{n-\frac{1}{2}}}{2} \times \boldsymbol{B}_0, \tag{11}$$

where $\Delta t$ is unit time step, $E$ and $P$ are assigned to integer time step, $H$ and $J$ are assigned to half integer time step, that is, $E$ and $P$ or $H$ and $J$ are calculated simultaneously. In particularly, it is necessary to solve the following matrix equation for (11) to obtain each component of $J$ in every time-step,

$$\begin{pmatrix} \frac{1}{\Delta t}+\frac{\gamma}{2} & -\frac{\varepsilon_0\omega_p^2}{q_e n_e}\frac{B_{z0}}{2} & \frac{\varepsilon_0\omega_p^2}{q_e n_e}\frac{B_{y0}}{2} \\ \frac{\varepsilon_0\omega_p^2}{q_e n_e}\frac{B_{y0}}{2} & \frac{1}{\Delta t}+\frac{\gamma}{2} & -\frac{\varepsilon_0\omega_p^2}{q_e n_e}\frac{B_{z0}}{2} \\ -\frac{\varepsilon_0\omega_p^2}{q_e n_e}\frac{B_{z0}}{2} & \frac{\varepsilon_0\omega_p^2}{q_e n_e}\frac{B_{y0}}{2} & \frac{1}{\Delta t}+\frac{\gamma}{2} \end{pmatrix} \begin{pmatrix} J_x^{n+\frac{1}{2}} \\ J_y^{n+\frac{1}{2}} \\ J_z^{n+\frac{1}{2}} \end{pmatrix} =$$

$$\begin{pmatrix} \left(-\frac{1}{\Delta t}+\frac{\gamma}{2}\right)J_x^{n-\frac{1}{2}} + \frac{\varepsilon_0\omega_p^2}{2q_e n_e}\left(J_y^{n-\frac{1}{2}}B_{z0} - J_z^{n-\frac{1}{2}}B_{y0}\right) + \varepsilon_0\omega_p^2 E_x^n - \omega_0^2 P_x^n \\ \left(-\frac{1}{\Delta t}+\frac{\gamma}{2}\right)J_y^{n-\frac{1}{2}} + \frac{\varepsilon_0\omega_p^2}{2q_e n_e}\left(J_z^{n-\frac{1}{2}}B_{x0} - J_x^{n-\frac{1}{2}}B_{z0}\right) + \varepsilon_0\omega_p^2 E_y^n - \omega_0^2 P_y^n \\ \left(-\frac{1}{\Delta t}+\frac{\gamma}{2}\right)J_z^{n-\frac{1}{2}} + \frac{\varepsilon_0\omega_p^2}{2q_e n_e}\left(J_x^{n-\frac{1}{2}}B_{y0} - J_y^{n-\frac{1}{2}}B_{x0}\right) + \varepsilon_0\omega_p^2 E_z^n - \omega_0^2 P_z^n \end{pmatrix}. \quad (12)$$

## 4. Numerical examples of millimeter-wave vortex propagation in magnetized plasma

Numerical model (*y-z* vertical cross-section) for simulation of the propagation of millimeter-wave vortex field in magnetized plasma (see Fig.2) is depicted in Fig.3. It is assumed that the hybrid mode millimeter-wave vortex with topological charge *l* is excited at the two wavelengths distance from the upstream edge of the corrugated waveguide with the radius $a = 40$ mm, and then the millimeter-wave vortex travels to the downstream vacuum region and illuminates the magnetized plasma. The whole grid size is taken to be $800 \times 800 \times 600$, and the unit grid size is 0.15 mm. Frequency and power of the excited vortex field are assumed to be 84 GHz and 1 MW, where $n_e$ and $|\boldsymbol{B_0}|$ are taken to be $1.1 \times 10^{20}$ m$^{-3}$ and 2 T, respectively. $\boldsymbol{B_0}$ is assumed to be oriented to x-direction.

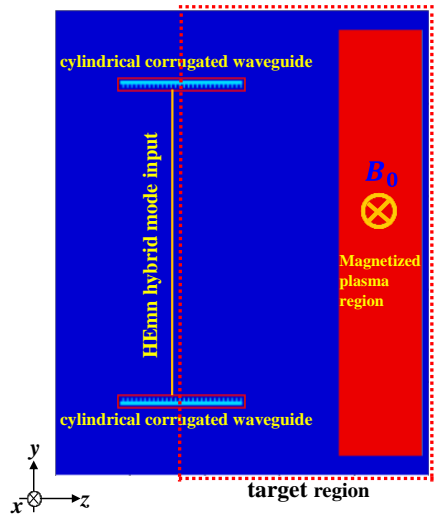

Fig.3 FDTD simulation of numerical model

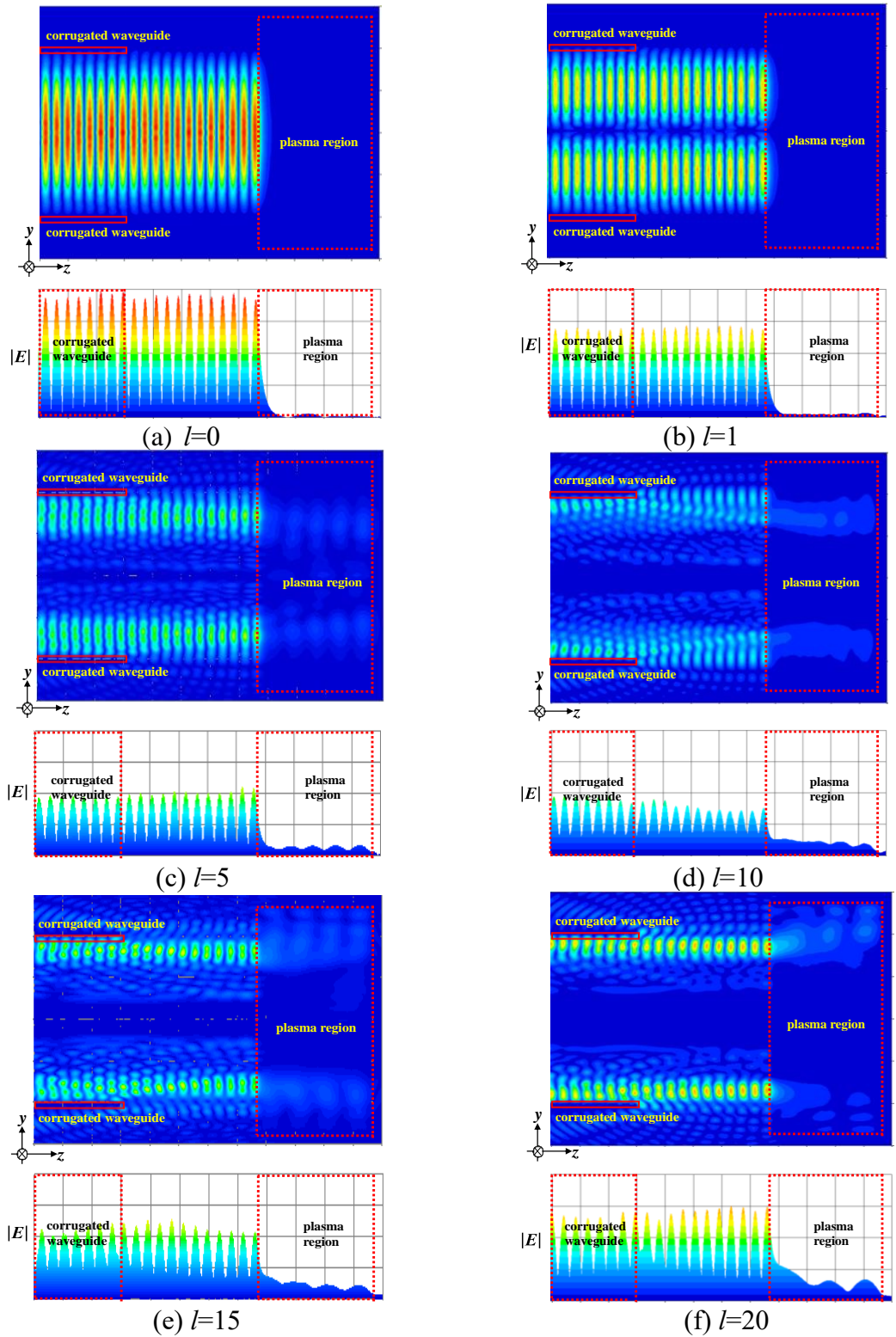

Fig.4 Electric field intensity in *y-z* plane

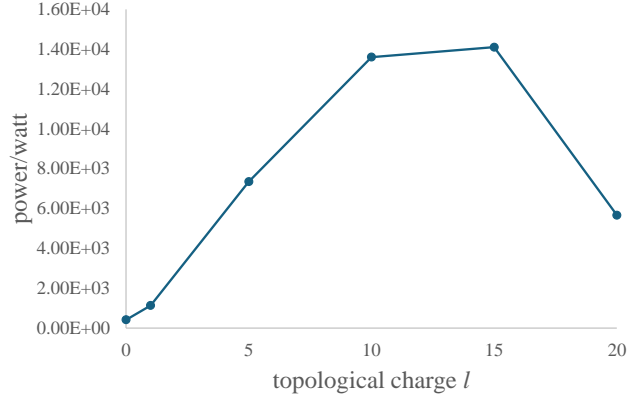

Fig.5 Propagation power of vortex fields in magnetized plasma for topological charge $l$

The electric field intensity distributions in y-z plane are depicted in Fig.4. Then Fig.4 (a), (b), (c), (d), (e) and (f) are for cases of $l$=0 (plane wave), 1, 5, 10, 15 and 20, respectively. It is assumed that the vortex fields are linearly polarized for x-direction, that corresponds to $\sigma_z = 0$ in (6). We can confirm that the millimeter-wave vortex of the hybrid mode of cylindrical corrugated waveguide can propagate in the magnetized plasma region in which the plane wave ($l$=0) can not propagate. That is, it is found that similar phenomena as in L-G mode vortex[12] can be existed in the hybrid mode vortex as well. In Fig.5, the propagation power of millimeter-wave vortex in the magnetized plasma is plotted for the topological charge $l$. It can be seen from the simulation results that the propagation power of millimeter-wave vortex in the magnetized plasma has the peak value at $l$=15, which is not predicted in the theoretical study for L-G mode vortex[12]. Indeed, the hybrid mode vortex is confined inside the corrugated waveguide and the field distribution is quite different from that of the L-G mode, in particularly, for a large value of the topological charge $l$. Accordingly, the propagation power of the hybrid mode vortex field may depend on the size of waveguide, which is different situation from that of the L-G mode vortex.

## 5. Conclusions

In this study, the propagation characteristic of the millimeter-wave vortex of the cylindrical corrugated waveguide hybrid mode in the magnetized plasma has been discussed by using 3D FDTD method. It is confirmed that the hybrid mode for millimeter-wave vortex field can propagate in the magnetized plasma region in which the normal plane wave is in cut-off condition. Then, it was also found that propagation power in the plasma region strongly depends on the topological charge $l$ and has peak value at $l$=15 in this study. We need further investigations on the dependence of the propagation power on the topological charge as a future work.

## Acknowledgments

The research was supported by KAKENHI (Nos. 21H04456, 22H05131, 23H04609, 22K18272, 23K03362), by the NINS program of Promoting Research by Networking among Institutions (01422301) by the NIFS Collaborative Research Programs (NIFS22KIIP003, NIFS24KIIT009, NIFS24KIPT013, NIFS22KIGS002, NIFS22KISS021).